%% file: Technical_Comment.tex
\documentclass[12pt]{article}
\usepackage{scicite}

\usepackage{times}
\usepackage{upgreek}
\usepackage[usenames, dvipsnames]{color}
\usepackage{graphicx}
\usepackage{gensymb}
\usepackage{subfiles}
\usepackage{rotating}
\usepackage{caption}
\usepackage{soul} 
\usepackage{hyperref}
\usepackage{listings}
\usepackage{upgreek}
\usepackage{footnote}

\newcommand{\ga}{\;\raisebox{-0.3ex}{\mbox{$\stackrel{>}{_\sim} \;$}}}

\definecolor{meatbrown}{rgb}{0.9, 0.72, 0.23}
\definecolor{palatinatepurple}{rgb}{0.41, 0.16, 0.38}
\definecolor{outerspace}{rgb}{0.5, 0.5, 0.5}
\definecolor{pear}{rgb}{0.82, 0.89, 0.19}
\definecolor{phthalogreen}{rgb}{0.07, 0.21, 0.14}
\definecolor{aqua}{rgb}{0.0, 1.0, 1.0}
\definecolor{aureolin}{rgb}{0.99, 0.93, 0.0}
\definecolor{babyblue}{rgb}{0.54, 0.81, 0.94}
\definecolor{brilliantrose}{rgb}{1.0, 0.33, 0.64}
\definecolor{sepia}{rgb}{0.44, 0.26, 0.08}
\definecolor{sunglow}{rgb}{1.0, 0.8, 0.2}
\definecolor{myred}{rgb}{0.9, 0.0, 0.0}
\definecolor{pinegreen}{rgb}{0.0, 0.5, 0.0}

\newenvironment{sciabstract}{%
\begin{quote} \bf}
{\end{quote}}

\title{Comment on ``A non-interacting low-mass black hole -- giant star binary system''} 
\author
{
Ed P. J. van den Heuvel$^{1\dagger}$,  
Thomas M. Tauris $^{2,3}$\\
\\
\normalsize{$^{1}$ Anton Pannekoek Institute for Astronomy, University of Amsterdam, Science~Park~904}\\
\normalsize{1098~XH Amsterdam, The Netherlands.}\\
\normalsize{$^{2}$ Aarhus Institute of Advanced Studies, Aarhus University, 8000~Aarhus~C, Denmark.}\\
\normalsize{$^{3}$ Department of Physics and Astronomy, Aarhus University, 8000~Aarhus~C, Denmark.}\\
\\
\normalsize{$^\dagger$To whom correspondence should be addressed: }
\normalsize{E-mail: E.P.J.vandenHeuvel@uva.nl}\\
 \\
 \\
 \\
 \\
 \\
 \\
 \\
 \\
Author version of Technical Comment published in\\Science on 8 May, 2020 (Vol. 368, Issue 6491, eaba3282).\\
For final publication, see https://doi.org/10.1126/science.aba3282 
}
 
\date{}

\begin{document} 

\input{science.tex}

\bibliography{scibib}

\bibliographystyle{Science}

\end{document}

%% file: science.tex
\baselineskip24pt

\maketitle 
\noindent

\clearpage
\begin{sciabstract} 
Thompson et al. (Reports, 1 November 2019, p. 637, Science) interpreted the unseen companion of the red giant star 2MASS J05215658+4359220 as most likely a black hole. We argue that if the red giant is about one solar mass, its companion can be a close binary consisting of two main-sequence stars. This would explain why no X-ray emission is detected from the system.
\end{sciabstract}

Thompson~et~al.~\cite{Thompson+2019} argued that the invisible companion in a 83-day orbit around the red giant star 2MASS~J05215658+4359220 is most likely a black hole, with a mass of $3.3_{-0.7}^{+2.8}$ solar masses ($M_\odot$). If this companion was a normal non-degenerate star, its light would have been detectable in the spectral energy distribution (SED) of the red giant, but was not. 

\bigskip
This approach assumes a normal single unseen companion star. However, in wide binaries the companion can itself be a closer binary composed of two normal stars with an orbital period of the order of a few days. Such hierarchical triple systems are quite common~\cite{rdl+13}. The luminosity $L$ of a main sequence star depends strongly on its mass $M$ ($L\propto M^{3.5}$), so a close binary consisting of two stars of equal mass has a luminosity about 6~times smaller than that of a single star with their combined mass. The light of such a binary companion might be undetectable in the SED of the red giant.

\bigskip
Thompson~et~al. argue that the red giant most likely has a mass of $3.2\;M_\odot$, but show that its high carbon-to-nitrogen ratio [C/N] is consistent with a lower mass of $1\;M_\odot$~\cite{Thompson+2019}. Spectroscopic determination of a red giant’s mass from model atmospheres can be uncertain by a factor~of~3~\cite{aum69}, which would allow for a red giant mass of about $1\;M_\odot$.

\bigskip
For the higher red giant mass, the companion is $3.3_{-0.7}^{+2.8}\;M_\odot$. If it were a close binary, the binary components would be $\sim\!1.65\;M_\odot$ each, which would have been detectable in the SED of the red~giant~\cite{Thompson+2019}. In this case a black hole companion is the only remaining possibility. 

\bigskip
In the alternative case of a $1\;M_\odot$ red giant, the mass of the unseen companion is $\ge 1.8\;M_\odot$~\cite{Thompson+2019}, so the companion could be a very close binary of two main-sequence stars of about $0.9\;M_\odot$, each, which would be of spectral type K0-4V with 0.44~solar~luminosities ($L_\odot$)~\cite{tag10}. Their combined luminosity would be less than 0.5\% of the $>200\;L_\odot$ luminosity of the red giant, which is undetectable.

\bigskip
Such a triple star system is expected to be dynamically stable if the ratio between the semi-major axis of the outer star and that of the inner binary is $\ga 3.0$~\cite{mik08}. Using Kepler’s third law, the upper limit on the orbital period of the inner binary is $\sim\! 20\;{\rm days}$. This would be consistent with a wide range of possible binaries, including very close systems of K-type main sequence stars. These are common, and often have a distant third companion~\cite{tok08}.

\bigskip
A black hole that accretes matter from the red giant’s wind would form an accretion disc, which might be detectable in X-ray emission. The physics of accretion of neutron stars and black holes is very similar. In both cases the accreting object becomes a strong X-ray source. There are thirteen known red-giant X-ray binaries, known as symbiotic X-ray binaries~\cite{csm+08,ykp19}. Six of them are regularly pulsating X-ray sources~\cite{ykp19}, showing that neutron stars accreting from the winds of red giants produce sources with X-ray luminosities, $L_X=10^{32}$ to $10^{36}\;{\rm erg\,s}^{-1}$. We expect similar values for accreting black holes. As Thompson~et~al.~\cite{Thompson+2019} show, using Bondi--Hoyle accretion~\cite{do73}, for typical red giant wind velocities, the black hole in the $3.2\;M_\odot$ giant case will capture 1\% of the total red-giant mass loss rate. 

\bigskip
The wind mass-loss rate of red giants is given by~\cite{rei75}:
\begin{equation}
     \dot{M}_{\rm w} = 4\times 10^{-13}\;\eta _R\,L\,R\,M^{-1}\qquad M_\odot\,{\rm yr}^{-1}
\end{equation}     
where $\eta _R$ is the efficiency parameter for wind mass loss and $L$, $R$ and $M$ are the luminosity, radius and mass of the giant in solar units. Observations of red giant wind mass~loss~\cite{mz15} show $\eta _R=0.477\pm 0.070$. Applying the nominal values for 2MASS~J05215658+4359220, $L= 331\;L_\odot$, $R=30\;R_\odot$~\cite{Thompson+2019}, and assuming the X-ray energy release $\Delta U = 0.1\;mc^2$ for a mass $m$ accreted onto a black hole, we obtain $L_X = 2.5\times 10^{34}\;{\rm erg\,s}^{-1}$ for a $3.2\;M_\odot$ giant with a $3.3\;M_\odot$ black hole. Thompson~et~al.~\cite{Thompson+2019} find a lower expected value of $L_X = 1.4\times 10^{33}\;{\rm erg\,s}^{-1}$. Such X-ray luminosities would be easily detectable, but no X-ray emission from the system is observed.

\bigskip
We consider a $1\;M_\odot$ red giant with a close-binary companion to be more consistent with the observational constraints on this system as it does not produce detectable light, nor X-ray emission, and the high [C/N] abundance ratio is normal for a $1\;M_\odot$ red~giant~\cite{Thompson+2019}.
A $3.2\;M_\odot$ red giant with a companion black hole would make this system unusual on three independent accounts: i) the [C/N] ratio of the giant would be unusually high for giants of this mass; ii) it has a black hole of unusual mass; and iii) an explanation must be found for the lack of X-ray emission.

\bigskip
We conclude that the unseen companion of 2MASS~J05215658 might not be a black hole.

%


\bigskip
\paragraph*{Acknowledgements}
We thank Mrs. A.M. van den Heuvel for drawing our attention to the paper by Thompson et al. (1). We thank Prof. Todd Thompson for kind and helpful explanations and efficient exchange of ideas on 2MASS J05215658+4359220. Funding: T.M.T. acknowledges an AIAS-COFUND Senior Fellowship funded by the European Union’s Horizon 2020 Research and Innovation Programme (grant no. 754513) and the Aarhus University Research Foundation. Author contributions: E.P.J.v.d.H. and T.M.T. interpreted the results, analyzed the system and wrote the manuscript together. Competing interests: We declare no competing interests. Data and materials availability: There are no new data in this Technical Comment. 

%

\clearpage